\documentclass[preprint,showpacs,groupedaddress,amsmath,amssymb]{revtex4}
\usepackage{graphicx}
\usepackage{dcolumn}
\usepackage{multirow}
\usepackage{amsfonts}

\newcommand{\beq}{\begin{equation}}
\newcommand{\eeq}{\end{equation}}
\newcommand{\bea}{\begin{align}}
\newcommand{\eea}{\end{align}}
\newcommand{\vepsi}{\epsilon}
\newcommand{\epsi}{\varepsilon}

\newcommand{\Cth}{\cos \theta}
\newcommand{\Sth}{\sin \theta}
\newcommand{\del}{\nabla}

\newcommand{\PI}{\overleftrightarrow{\mathbf{\Pi}}}

\newcommand{\rhat}{\hat{r}}
\newcommand{\thetahat}{\hat{\theta}}
\newcommand{\phihat}{\hat{\phi}}

\newcommand{\mbf}[1]{\mathbf{#1}}

\newcommand{\Bpol}{B_\theta}
\newcommand{\Btor}{B_\phi}
\newcommand{\Vpol}{V_\theta}

\newcommand{\Vpols}{V_{\theta\, s}}
\newcommand{\Vtors}{V_{\phi\, s}}
\newcommand{\divr}{\nabla \cdot}
\newcommand{\curl}{\nabla \times}

\newcommand{\ddt}[1]{\frac{\partial\, {#1}}{\partial t}}

\makeatletter
\def\Tbar{\mathchoice
   {\TTbar\displaystyle\textstyle{-}}%
   {\TTbar\textstyle\scriptstyle{-}}%
   {\TTbar\scriptstyle\scriptscriptstyle{-}}%
   {\TTbar\scriptscriptstyle\scriptscriptstyle{-}}%
   \!T}
\def\TTbar#1#2#3{{\setbox0=\hbox{$#1{#2#3}{\mathrm{T}}$}
     \raise2\p@\vbox{\hbox{$#2#3$}}\kern-.35\wd0}}
\makeatother

\begin{document}

\title{On ``the plasma approximation''}


\author{Robert W. Johnson}
\email[]{rob.johnson@gatech.edu}
\affiliation{Atlanta, GA 30238, USA}


\date{\today}

\begin{abstract}
``The plasma approximation'' is, in the words of Pauli, ``not even wrong,'' as it expresses a disbelief in the symmetries of the underlying gauge field theory.  The electrostatic field in question is divergenceful yet has no sources or sinks, thus breaking Lorentz covariance explicitly, and has no place in a self-consistent theory.  Utilizing it leads to inconsistencies in the equation of motion and prevents a proper, field-theoretic treatment of condensed matter in the plasma state.  By exposing the fallacy of replacing one of the field equations with one of the Bianchi identities, the applicability of Gauss' law to plasma physics is assured.
\end{abstract}


\maketitle


All approximations are wrong; the question is, how wrong?  ``The plasma approximation'' which is assumed applicable to low-frequency and steady-state phenomenon, popular among plasma scientists in various fields~\cite{chen-84,fshu} and tracing its roots beyond Braginskii~\cite{brag-1965,roseclark}, is, in the words of Pauli, ``not even wrong,'' as it expresses a disbelief in the symmetries of the underlying gauge field theory.  (Many authors~\cite{jabbook-04,benzbook-02} use the term to mean what is better called ``the plasma condition''~\cite{NWS-2000}, {\it ie} the number of electrons in a Debye sphere is large enough to effect charge shielding, a situation akin to the vacuum polarization found in quantum electrodynamics~\cite{halzenmartin}---that is not the issue herein.  What is at issue here is the presupposition of a diverging electrostatic field on scales much larger than the Debye length in the absence of an explicit supporting charge density.)  The electrostatic field in question is divergenceful yet has no sources or sinks.  Utilizing it leads to inconsistencies in the equation of motion and prevents a proper, field-theoretic treatment of condensed matter in the plasma state.  As Griffiths says~\cite{griffiths-89}, ``like Gauss' law, Ampere's law [with Maxwell's correction] is always {\it true}.''  The reason is because these equations ``may be regarded as a four-dimensional version of Poisson's equation.''  Lorentz covariance implies a symmetry in the inhomogeneous of Maxwell's equations which must be respected, as decoupling electrostatic fields from source charges is tantamount to decoupling magnetostatic fields from source currents.  We reiterate the development of classical and quantum field theory~\cite{griffiths-89,davis70,ramond-FTM90,ryder-qft,mandlshaw} and insist on its applicability to all media, including magnetized plasma.

Let us first be pedantic with our terminology, as only formal definitions equate with formal mathematics.  The term ``quasineutral'' relates to ``the plasma condition'' defined above, referring to the overwhelming tendency of electrons within a plasma to shift their positions in response to the electrostatic potential of the ion density so as to attenuate exponentially the Coulomb field, and is often taken as the definition of ``the plasma approximation.''  However, the traditional term ``neutral'' already encompasses the implications of quasineutrality, as no medium (except that of neutron stars~\cite{acphillips}) remains neutral on scales sufficiently small to resolve isolated charges.  Within neutral media, the microscopic field fluctuates wildly on the particle scale but averages out as the differential volume element grows~\cite{griffiths-89,feynmanlecs}.  In contrast, as Chen succinctly states~\cite{chen-84}, ``the plasma approximation is almost the same as the condition of quasineutrality discussed earlier but has a more exact meaning [and] is a mathematical shortcut that one can use even for wave motions.''  Specifically, ``it is usually possible to assume $n_i = n_e$ and $\divr \mbf{E} \neq 0$ at the same time.''  Indeed, most modern definitions of the plasma magnetohydrodynamic equations~\cite{chen-84,dendybook-93,dinkbook-05,staceybook05,kivel95,spsbds03,mbk04,goldruth95} include only three-fourths of Maxwell's laws.  Regardless of the name one gives the procedure, the {\it idea} that Gauss' law is inapplicable within a plasma runs contrary to well over a century's worth of experimental and theoretical analysis~\cite{maxwell-1864} and leads one into a host of difficulties in determining the physics of the plasma state.

The following draws heavily (nearly {\it verbatim}) from Mandl and Shaw's {\it Quantum Field Theory}~\cite{mandlshaw}.  Starting with Maxwell's microscopic equations, \beq
\divr \mbf{E} = \frac{\rho}{\vepsi_0}\;,\; \curl \mbf{B} = \mu_0 \mbf{J} + \mu_0 \vepsi_0 \ddt{}\mbf{E}\;, \eeq
\beq \divr \mbf{B} = 0\;,\; \curl \mbf{E} = -\ddt{}\mbf{B}\;,
\eeq write the antisymmetric field tensor \beq
F^{\mu\,\nu} \equiv \left[ \begin{array}{cccc} 0 & E_x/c & E_y/c & E_z/c \\ -E_x/c & 0 & B_z & -B_y \\ -E_y/c & -B_z & 0 & B_x \\ -E_z/c & B_y & -B_x & 0 \end{array} \right]\;,
\eeq where $\mu,\nu\in\{0,1,2,3\}$ give the row and column, respectively and $c\equiv1/\sqrt{\vepsi_0 \mu_0}$ is the speed of light in vacuum.  The metric tensor is $g_{0\,0}=-g_{j\,j}=+1$ for $j\in\{1,2,3\}$ and $g_{\mu\,\nu}=0$ otherwise, and $\partial_\mu \equiv \partial/\partial x^\mu$.  With the charge-current density $s^\mu(x) \equiv (c\rho(x),\mbf{J}(x))$, Maxwell's equations become
\beq \label{eqn:max1} \partial_\nu F^{\mu\,\nu}(x) = \mu_0 s^\mu(x)\;, \eeq
\beq \label{eqn:max2} \partial^\lambda F^{\mu\,\nu}(x) + \partial^\mu F^{\nu\,\lambda}(x) +\partial^\nu F^{\lambda\,\mu}(x) = 0\;. \eeq
Antisymmetry in $F^{\mu\,\nu}$ gives $\partial_\mu s^\mu(x)=0$, conservation of the current coupled to the electromagnetic field.  In terms of the four-vector potential $A^\mu(x)\equiv(\Phi/c,\mbf{A})$, the field tensor is expressed as \beq 
F^{\mu\,\nu}(x) = \partial^{\,\nu} A^\mu(x) - \partial^{\,\mu} A^\nu(x)\;,
\eeq which is equivalent to \beq
\mbf{B}=\curl\mbf{A}\;,\; \mbf{E}=-\del\Phi-\ddt{}\mbf{A}\;.
\eeq In terms of $A^\mu(x)$, Equations~(\ref{eqn:max2}), known as the Bianchi identities, are satisfied identically, and Equations~(\ref{eqn:max1}), known as ``the'' field equations, read \beq \label{eqn:max3}
\partial_\nu \partial^\nu (A^\mu(x)) - \partial^\mu(\partial_\nu A^\nu(x)) = \mu_0 s^\mu(x)\;,
\eeq which are Lorentz covariant and gauge invariant.  Consistency upon quantization requires $\partial_\nu A^\nu(x)=0$, which is recognized as the Lorentz condition, achievable from any given configuration by a gauge transformation.  Incorporating ``the plasma approximation'' amounts simply to replacing the $=$ signs with $\neq$ signs in Equations~(\ref{eqn:max1}) and (\ref{eqn:max3}).  Thus, one sees that Gauss' law is not an isolated, arbitrary equation, being part of the unification of electrodynamics brought on by Maxwell~\cite{maxwell-1864}, and that its violation amounts to decoupling magnetostatic fields from source currents as well as electrostatic fields from source charges, thus breaking the symmetries of the underlying gauge field theory, hence is incompatible with the laws of physics.

Another way to look at the problem is to consider the manner in which students of plasma physics are introduced to the concept (starting at the undergraduate level!) by following the arguments presented by Chen~\cite{chen-84} and Dougherty~\cite{dendybook-93} (which in no way reflects a criticism of these particular authors, and in fact compliments them on such a clear elucidation of the topic at hand):  Using continuity, take the divergence of Maxwell's curl equations to show that his divergence equations are ``redundant'' and may be dropped from the analysis.  However, from the development above, we see that the natural, physical division of Maxwell's laws is not into the ``curl'' and ``divergence'' equations but rather into the homogeneous and inhomogeneous equations, and that dropping the electric field's divergence equation in favor of its curl equation amounts to the replacement of an information bearing field equation with one of the Bianchi identities, thus literally leaving useful information out of the subsequent analysis and hence losing touch with the real physics.  At the classical level, simply rewriting the fields in terms of the four-potential obviates the question of redundant equations.  The subsequent justification of the validity of ``the plasma approximation'' given by Chen in the context of the linear theory of ion acoustic waves actually only justifies the neglect of perturbations to the ion density down to scales on the order of the Debye length, not the neglect of Gauss' law, as its identical twin Poisson's equation is used throughout the analysis.

Let us examine a recent appearance~\cite{npg-14-49-2007} and discussion~\cite{npg-14-543-2007,npg-14-545-2007} of the issue in the geophysical literature as it pertains to ``large amplitude stationary structures in general plasmas'' in a ``static magnetic field''.  Here, Verheest finds that demanding ``strict charge neutrality [on a differential volume element, one must assume from the fluid formalism] has indicated that solitons and oscillitons cannot exist in electron-ion plasmas,'' rather than the suggestion by McKenzie, {\it et al}~\cite{npg-12-425-2005}, ``that fully relativistic effects may prevent the formation of oscilliton-like structures.''  From the discussion above, we see that these arguments are not unrelated, nor are they in conflict, as only by respecting Gauss' law does Maxwell's theory remain Lorentz covariant.  The statement by McKenzie, {\it et al}, in their Comment, ``that a question of using Poisson's equation versus the quasineutrality condition is not a new one [and] was discussed and clarified 40 years ago in the classical review paper by Braginskii~\cite{brag-1965},'' is not accurate.  The relevant discussion surrounding Equation~(6.13) thereof is declaratory and unreferenced, hardly ``clarified,'' yet has been taken as gospel by subsequent generations of plasma physicists in a variety of applications.  Earlier treatments~\cite{roseclark} focus on the argument that one cannot get more than one piece of information out of Poisson's equation, and as Equations~(\ref{eqn:max3}) show, one does not, as the electrostatic potential and the supporting charge density are intimately related.

Turning now to a popular application of ``the plasma approximation'' to the evaluation of the electrostatic field in a tokamak discharge at equilibrium within the prescripts of the neoclassical model~\cite{solomonetal-pop-2006,stacey-cpp06,frc-pop-2006} (some geometric nomenclature is given in the caption of Figure~1, and these technical arguments serve merely to support, not to define, the main contention of this article), we note that using an impurity ion species' radial equation of motion to determine a non-vanishing, non-poloidally varying radial electrostatic field is at best an indirect measurement subject to the assumptions of the underlying model.  (The direct measurement of the radial electric field accomplished by motional Stark effect diagnostics~\cite{holcomb-10E506,cortes-1596} viewing the $D_\alpha$ line emission of injected neutrals excited by collisions with plasma particles is biased towards observations along the horizontal midplane as a consequence of port placement hence does not necessarily reveal the true radial field profile in the $(Z,R)$ plane; neither does it distinguish between static and dynamic electric fields.)  The neoclassical prescription to use an equation of motion, rather than Gauss' law, to determine the radial electrostatic field, and then to turn around and use Poisson's equation to determine the charge density, is tautologically inconsistent, as the implication of $\divr \mbf{E} \neq \rho / \vepsi_0$ must be that $-\del^2 \Phi \neq \rho / \vepsi_0$.  In his development of the fluid equation of motion from the first moment of the Vlasov equation~\cite{dendybook-93}, Elliot states that ``the second integral [in Equation (2.11.10) thereof] is also zero, because $\mbf{E}$ is not a function of'' the particle velocities.  Thus, neither is $\mbf{E}$ a function of their average, the fluid velocity $\mbf{V}$.  Rather, it should be considered as an {\it input} to the equation of motion, which is a differential equation to be solved for $\mbf{V}$.  The defining relations for $\mbf{E}$ are, naturally, Maxwell's, and the use of the fluid equation of motion to {\it determine} $\mbf{E}$ is inconsistent with its own derivation.

Nonetheless, the fluid equation of motion, Equation~(5) in Reference~\cite{solomonetal-pop-2006} giving its radial component in the neoclassical model, must hold even when a velocity component vanishes, as does the radial component at equilibrium, leaving its interpretation in a bit of a conundrum:  if the defining properties of the plasma state include the requirement of charge shielding, then the net charge on a differential volume element, which must be much larger than the Debye length yet smaller than any other scales of consideration for the averaging inherent in the definition of continuum quantities such as density, temperature, {\it etc}, to be formally valid, necessarily must vanish, as does any emitted electric flux.  Taken at face value, the electrostatic potential, Figure~2, calculated from the radial field presented by Solomon, {\it et al}, in Figure~12(c) of Reference~\cite{solomonetal-pop-2006}, is not of harmonic form~\cite{simmons-91,flanigan,BLT-92,wolf-lapeqn}, Figure~3, thus it implies a shifting of the electron density profile relative to the ions on the order of $O(\pm10^{13}/m^3)$, Figure~4, which does not sound like much when the absolute density is on the order of $O(10^{19}/m^3)$, yet produces a non-cancellation of terms which usually vanish in the quasineutral treatment, $n_e = \sum_{ions} z_i n_i$ for species charge $e_s\equiv z_s e$.  The first inconsistency arises in the supposition of no poloidal variation to the radial field, as the radial equation of motion retains a poloidal dependence when the intrinsic variation of quantities induced by the geometry is considered~\cite{mingagrees}, given in the large aspect ratio limit, $\epsi \equiv r/R_0 \ll 1$, by \beq
\mbf{B}=\mbf{B}^0/(1+\epsi\Cth)\;,\; \Vpols=\Vpols^0/(1+\epsi\Cth)\;,\; \Vtors=\Vtors^0(1+\epsi\Cth)\;,\; n_s = n_s^0\;,
\eeq where $A^0\equiv\langle A\rangle \equiv \oint d\theta (1+\epsi\Cth) A/2\pi$ is the flux-surface average of $A$.  Using these values, \beq
E_r(\theta) = p'^{\,0}_s/e_s n^0_s + \Vtors^0 \Bpol^0 - \Vpols^0 \Btor^0 /(1+\epsi\Cth)^2\;,
\eeq for species pressure $p'_s \equiv \partial n_s \Tbar_s / \partial r$, thus $E_r(\theta) \not\equiv E_r^0$.  (A further complication occurs if the species temperature $\Tbar_s \equiv k T_s$ retains a poloidal dependence.)  Reconsidering that equation in light of a non-vanishing charge density, $\sum_s e_s n_s \neq 0$, we sum over species to obtain \beq
E_r \sum_s e_s n_s = \sum_s \left[p'_s + e_s n_s \left( \Vtors \Bpol - \Vpols \Btor \right) \right] \equiv 0 \;,
\eeq where the l.h.s. may be interpreted physically as proportional to $1 \times 1$ or $0 \times 0$ or neoclassically as $1 \times 0$.  Noting that the r.h.s. is the radial component of the equilibrium equation $\del p = \mbf{J} \times \mbf{B}$ lets one identify it as zero, leading to inconsistency for a non-vanishing charge density.  The alternative interpretation is to treat the plasma as a conducting fluid which is {\it neutral} on scales down to the Debye length and to reexamine the convective and viscous terms usually neglected, to wit, \beq
n_s m_s\left[\left(\mbf{V}_s \cdot \del \right)\mbf{V}_s\right]_r + \rhat \cdot \del \cdot \PI_s = n_s e_s \left(\Vpols \Btor - \Vtors \Bpol \right) - p'_s\;,
\eeq
where we note that the convective term has a contribution $\sim n m\Vpol^2/r$ which survives incompressibility and the flux surface average.  Thus, when one measures a species' net toroidal and poloidal velocities (in addition to the usual density, temperature, and magnetic field measurements), one might actually be determining that species' net radial viscous force.

From our demonstration of the incompatibility of ``the plasma approximation'' with the laws of physics as expressed in the language of field theory, as well as our identification of some of the inconsistencies that arise from its implementation, what are we to conclude?  Unfortunately and with some regret, we must conclude that much of the last half-century's worth of analysis on condensed matter in the plasma state within the fluid formalism of magnetohydrodynamics is gravely in error, being based upon beliefs and equations incompatible with the laws of Nature.  By neglecting Gauss' law, one cuts oneself off from the rich, powerful, and beautiful mathematics of field theory.  Leslie C. Woods~\cite{woodsbook-06,woodsobit} was correct in his assessment of the state of fusion plasma research and the need to reevaluate some of the established beliefs long-cherished in the field.  Perhaps with a more reactionary approach to plasma physics and fusion engineering, one which respects {\it all} of Maxwell's laws, the 21st century will see the fulfillment of ``the promise of fusion energy~\cite{jfe18-85}.''



\begin{figure}%
\includegraphics[scale=.5]{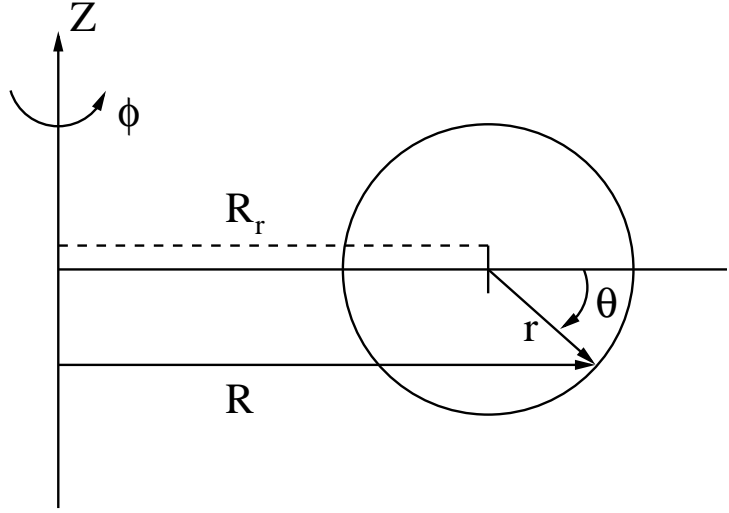}
\caption{Cylindrical coordinates $(Z,R,\phi)$ relate to tokamak coordinates $(r,\theta,\phi)$ via $Z=-r\Sth$ and $R=R_0+r\Cth$ in the concentric circular approximation, where $\rhat$, $\thetahat$, and $\phihat$ give the radial, poloidal, and toroidal directions, respectively.  The outermost minor radius of the confined plasma, given in meters by $a$, defines the normalized minor radius $\rho=r/a$, and $R_a$ is its centroid.  The magnetic field $\mbf{B}$ and current density $\mbf{J}$ lie in isobaric surfaces given by $\del p = \mbf{J}\times\mbf{B}$ for a stationary equilibrium, defining the ``flux surface'' at radius $r$.  In general, the nested flux surfaces are neither circular nor concentric.}
\end{figure}

\begin{figure}%
\includegraphics[scale=.5]{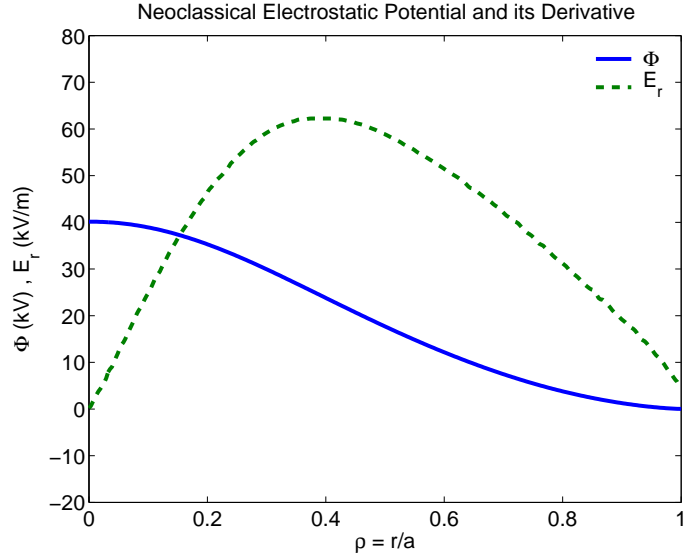}
\caption{Neoclassical electrostatic potential $\Phi$ and its derivative $E_r\equiv -\partial \Phi / \partial r$ {\it vs} normalized minor radius $\rho$.  The potential is evaluated numerically from the radial field given in Figure~12(c) of Reference~\cite{solomonetal-pop-2006}.}
\end{figure}

\begin{figure}%
\includegraphics[scale=.5]{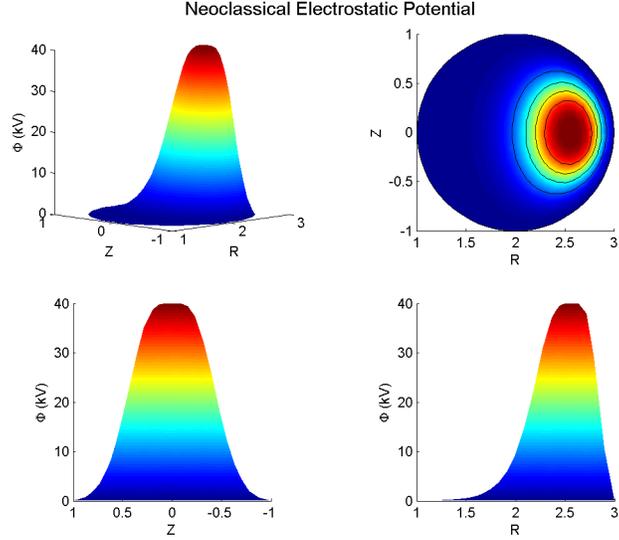}
\caption{(Color online.)  Neoclassical electrostatic potential $\Phi$ in the $(Z,R)$ plane, assuming no poloidal dependence, painted onto a generic equilibrium.  Note that the potential is not of harmonic form, as the value at the center is not equal to the average of the values around the boundary, hence requires a supporting charge density.}
\end{figure}

\begin{figure}%
\includegraphics[scale=.5]{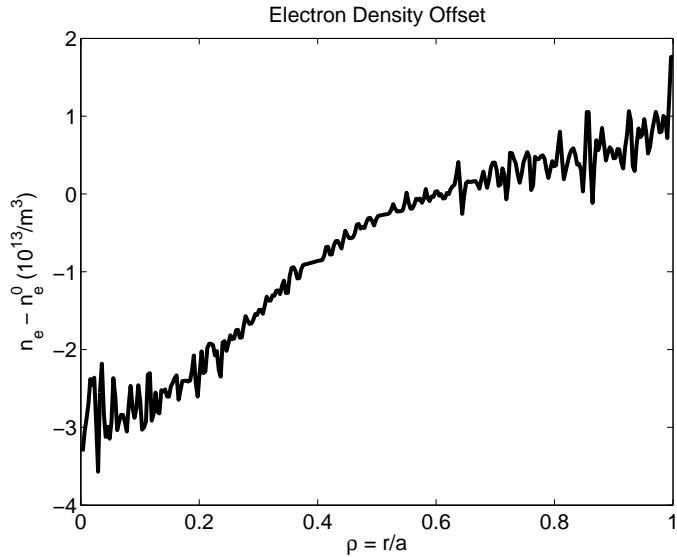}
\caption{Flux-surface averaged electron density offset given by Poisson's equation required to support the neoclassical electrostatic potential in units of $10^{13}/m^3$.}
\end{figure}

\clearpage



%


\begin{thebibliography}{40}
\expandafter\ifx\csname natexlab\endcsname\relax\def\natexlab#1{#1}\fi
\expandafter\ifx\csname bibnamefont\endcsname\relax
  \def\bibnamefont#1{#1}\fi
\expandafter\ifx\csname bibfnamefont\endcsname\relax
  \def\bibfnamefont#1{#1}\fi
\expandafter\ifx\csname citenamefont\endcsname\relax
  \def\citenamefont#1{#1}\fi
\expandafter\ifx\csname url\endcsname\relax
  \def\url#1{\texttt{#1}}\fi
\expandafter\ifx\csname urlprefix\endcsname\relax\def\urlprefix{URL }\fi
\providecommand{\bibinfo}[2]{#2}
\providecommand{\eprint}[2][]{\url{#2}}

\bibitem[{\citenamefont{{Chen}}(1984)}]{chen-84}
\bibinfo{author}{\bibfnamefont{F.~F.} \bibnamefont{{Chen}}},
  \emph{\bibinfo{title}{Introduction to Plasma Physics and Controlled Fusion}}
  (\bibinfo{publisher}{Springer}, \bibinfo{year}{1984}), ISBN
  \bibinfo{isbn}{0306413329}.

\bibitem[{\citenamefont{{Shu}}(1991)}]{fshu}
\bibinfo{author}{\bibfnamefont{F.~H.} \bibnamefont{{Shu}}},
  \emph{\bibinfo{title}{The Physics of Astrophysics}}
  (\bibinfo{publisher}{University Science Books}, \bibinfo{year}{1991}).

\bibitem[{\citenamefont{{Braginskii}}(1965)}]{brag-1965}
\bibinfo{author}{\bibfnamefont{S.~I.} \bibnamefont{{Braginskii}}}, in
  \emph{\bibinfo{booktitle}{Review of Plasma Physics}}, edited by
  \bibinfo{editor}{\bibfnamefont{M.}~\bibnamefont{Leontovich}}
  (\bibinfo{publisher}{Consultants Bureau}, \bibinfo{address}{New York,
  U.S.A.}, \bibinfo{year}{1965}), vol.~\bibinfo{volume}{1} of
  \emph{\bibinfo{series}{Review of Plasma Physics}}, pp.
  \bibinfo{pages}{201--311}.

\bibitem[{\citenamefont{{Rose} and C.}(1961)}]{roseclark}
\bibinfo{author}{\bibfnamefont{D.~J.} \bibnamefont{{Rose}}} \bibnamefont{and}
  \bibinfo{author}{\bibfnamefont{C.}~\bibnamefont{C.}},
  \emph{\bibinfo{title}{Plasmas and Controlled Fusion}}
  (\bibinfo{publisher}{M.I.T. Press}, \bibinfo{year}{1961}).

\bibitem[{\citenamefont{{Bittencourt}}(2004)}]{jabbook-04}
\bibinfo{author}{\bibfnamefont{J.~A.} \bibnamefont{{Bittencourt}}},
  \emph{\bibinfo{title}{Fundamentals of Plasma Physics}}
  (\bibinfo{publisher}{Springer}, \bibinfo{year}{2004}).

\bibitem[{\citenamefont{{Benz}}(2002)}]{benzbook-02}
\bibinfo{author}{\bibfnamefont{A.~O.} \bibnamefont{{Benz}}},
  \emph{\bibinfo{title}{Plasma Astrophysics: Kinetic Processes in Solar and
  Stellar Coronae}} (\bibinfo{publisher}{Springer}, \bibinfo{year}{2002}).

\bibitem[{\citenamefont{{Nishikawa} and {Wakatani}}(2000)}]{NWS-2000}
\bibinfo{author}{\bibfnamefont{K.}~\bibnamefont{{Nishikawa}}} \bibnamefont{and}
  \bibinfo{author}{\bibfnamefont{M.}~\bibnamefont{{Wakatani}}},
  \emph{\bibinfo{title}{Plasma Physics: Basic Theory with Fusion Applications}}
  (\bibinfo{publisher}{Springer}, \bibinfo{year}{2000}).

\bibitem[{\citenamefont{{Halzen} and {Martin}}(1985)}]{halzenmartin}
\bibinfo{author}{\bibfnamefont{F.}~\bibnamefont{{Halzen}}} \bibnamefont{and}
  \bibinfo{author}{\bibfnamefont{A.~D.} \bibnamefont{{Martin}}},
  \emph{\bibinfo{title}{Quarks and Leptons}} (\bibinfo{publisher}{Wiley},
  \bibinfo{year}{1985}).

\bibitem[{\citenamefont{{Griffiths}}(1989)}]{griffiths-89}
\bibinfo{author}{\bibfnamefont{D.}~\bibnamefont{{Griffiths}}},
  \emph{\bibinfo{title}{Introduction to Electrodynamics}}
  (\bibinfo{publisher}{{Prentice-Hall, Inc.}}, \bibinfo{address}{Englewood
  Cliffs, NJ, USA}, \bibinfo{year}{1989}), \bibinfo{edition}{2nd} ed., ISBN
  \bibinfo{isbn}{0-13-481367-7}.

\bibitem[{\citenamefont{Ramond}(1990)}]{ramond-FTM90}
\bibinfo{author}{\bibfnamefont{P.}~\bibnamefont{Ramond}},
  \emph{\bibinfo{title}{Field Theory: {A} Modern Primer}}
  (\bibinfo{publisher}{Addison-Wesley Publishing Company},
  \bibinfo{year}{1990}), \bibinfo{edition}{2nd} ed., ISBN
  \bibinfo{isbn}{0-201-54611-6}.

\bibitem[{\citenamefont{{Ryder}}(1985)}]{ryder-qft}
\bibinfo{author}{\bibfnamefont{L.~H.} \bibnamefont{{Ryder}}},
  \emph{\bibinfo{title}{Quantum Field Theory}} (\bibinfo{publisher}{Cambridge
  University Press}, \bibinfo{year}{1985}).

\bibitem[{\citenamefont{{Mandl} and {Shaw}}(1993)}]{mandlshaw}
\bibinfo{author}{\bibfnamefont{F.}~\bibnamefont{{Mandl}}} \bibnamefont{and}
  \bibinfo{author}{\bibfnamefont{G.}~\bibnamefont{{Shaw}}},
  \emph{\bibinfo{title}{Quantum Field Theory, Revised Edition}}
  (\bibinfo{publisher}{{John Wiley \& Sons Inc}}, \bibinfo{year}{1993}), ISBN
  \bibinfo{isbn}{0471941867}.

\bibitem[{\citenamefont{{Davis}}(1970)}]{davis70}
\bibinfo{author}{\bibfnamefont{W.~R.} \bibnamefont{{Davis}}},
  \emph{\bibinfo{title}{Classical Fields, Particles, and the Theory of
  Relativity}} (\bibinfo{publisher}{Gordon and Breach Science Publishers},
  \bibinfo{year}{1970}).

\bibitem[{\citenamefont{{Phillips}}(1994)}]{acphillips}
\bibinfo{author}{\bibfnamefont{A.~C.} \bibnamefont{{Phillips}}},
  \emph{\bibinfo{title}{The Physics of Stars}} (\bibinfo{publisher}{John Wiley
  and Sons}, \bibinfo{year}{1994}).

\bibitem[{\citenamefont{Feynman et~al.}(1963)\citenamefont{Feynman, Leighton,
  and Sands}}]{feynmanlecs}
\bibinfo{author}{\bibfnamefont{R.}~\bibnamefont{Feynman}},
  \bibinfo{author}{\bibfnamefont{R.~B.} \bibnamefont{Leighton}},
  \bibnamefont{and} \bibinfo{author}{\bibfnamefont{M.~L.} \bibnamefont{Sands}},
  \emph{\bibinfo{title}{The Feynman Lectures on Physics}}
  (\bibinfo{publisher}{Addison-Wesley}, \bibinfo{year}{1963}), \bibinfo{note}{3
  volumes}.

\bibitem[{\citenamefont{{Dendy}}(1993)}]{dendybook-93}
\bibinfo{author}{\bibfnamefont{R.}~\bibnamefont{{Dendy}}},
  \emph{\bibinfo{title}{Plasma Physics: an Introductory Course}}
  (\bibinfo{publisher}{Cambridge University Press},
  \bibinfo{address}{Cambridge, England}, \bibinfo{year}{1993}).

\bibitem[{\citenamefont{Dinklage et~al.}(2005)\citenamefont{Dinklage, Klinger,
  Marx, and Schweikhard}}]{dinkbook-05}
\bibinfo{editor}{\bibfnamefont{A.}~\bibnamefont{Dinklage}},
  \bibinfo{editor}{\bibfnamefont{T.}~\bibnamefont{Klinger}},
  \bibinfo{editor}{\bibfnamefont{G.}~\bibnamefont{Marx}}, \bibnamefont{and}
  \bibinfo{editor}{\bibfnamefont{L.}~\bibnamefont{Schweikhard}}, eds.,
  \emph{\bibinfo{title}{Plasma Physics: Confinement, Transport and Collective
  Effects}} (\bibinfo{publisher}{Springer}, \bibinfo{year}{2005}).

\bibitem[{\citenamefont{Stacey}(2005)}]{staceybook05}
\bibinfo{author}{\bibfnamefont{W.~M.} \bibnamefont{Stacey}},
  \emph{\bibinfo{title}{Fusion Plasma Physics}}
  (\bibinfo{publisher}{Wiley-VCH}, \bibinfo{year}{2005}).

\bibitem[{\citenamefont{Kivelson and Russell}(1995)}]{kivel95}
\bibinfo{author}{\bibfnamefont{M.~G.} \bibnamefont{Kivelson}} \bibnamefont{and}
  \bibinfo{author}{\bibfnamefont{C.~T.} \bibnamefont{Russell}},
  \emph{\bibinfo{title}{Introduction to Space Physics}}
  (\bibinfo{publisher}{Cambridge University Press}, \bibinfo{year}{1995}).

\bibitem[{\citenamefont{B\"{u}chner et~al.}(2003)\citenamefont{B\"{u}chner,
  Dum, and Scholer}}]{spsbds03}
\bibinfo{editor}{\bibfnamefont{J.}~\bibnamefont{B\"{u}chner}},
  \bibinfo{editor}{\bibfnamefont{C.~T.} \bibnamefont{Dum}}, \bibnamefont{and}
  \bibinfo{editor}{\bibfnamefont{M.}~\bibnamefont{Scholer}}, eds.,
  \emph{\bibinfo{title}{Space Plasma Simulation}}
  (\bibinfo{publisher}{Springer}, \bibinfo{year}{2003}).

\bibitem[{\citenamefont{Kallenrode}(2004)}]{mbk04}
\bibinfo{editor}{\bibfnamefont{M.-B.} \bibnamefont{Kallenrode}}, ed.,
  \emph{\bibinfo{title}{Space Physics: An Introduction to Plasmas and Particles
  in the Heliosphere and Magnetospheres}} (\bibinfo{publisher}{Springer},
  \bibinfo{year}{2004}).

\bibitem[{\citenamefont{Goldston and Rutherford}(1995)}]{goldruth95}
\bibinfo{author}{\bibfnamefont{R.~J.} \bibnamefont{Goldston}} \bibnamefont{and}
  \bibinfo{author}{\bibfnamefont{P.~H.} \bibnamefont{Rutherford}},
  \emph{\bibinfo{title}{Introduction to Plasma Physics}}
  (\bibinfo{publisher}{CRC Press}, \bibinfo{year}{1995}).

\bibitem[{\citenamefont{{Maxwell}}(1864)}]{maxwell-1864}
\bibinfo{author}{\bibfnamefont{J.~C.} \bibnamefont{{Maxwell}}},
  \bibinfo{journal}{Royal Society Transactions} \textbf{\bibinfo{volume}{155}}
  (\bibinfo{year}{1864}).

\bibitem[{\citenamefont{Verheest}(2007{\natexlab{a}})}]{npg-14-49-2007}
\bibinfo{author}{\bibfnamefont{F.}~\bibnamefont{Verheest}},
  \bibinfo{journal}{Nonlinear Processes in Geophysics}
  \textbf{\bibinfo{volume}{14}}, \bibinfo{pages}{49}
  (\bibinfo{year}{2007}{\natexlab{a}}), ISSN \bibinfo{issn}{1023-5809},
  \urlprefix\url{http://www.nonlin-processes-geophys.net/14/49/2007/}.

\bibitem[{\citenamefont{McKenzie et~al.}(2007)\citenamefont{McKenzie, Dubinin,
  and Sauer}}]{npg-14-543-2007}
\bibinfo{author}{\bibfnamefont{J.~F.} \bibnamefont{McKenzie}},
  \bibinfo{author}{\bibfnamefont{E.}~\bibnamefont{Dubinin}}, \bibnamefont{and}
  \bibinfo{author}{\bibfnamefont{K.}~\bibnamefont{Sauer}},
  \bibinfo{journal}{Nonlinear Processes in Geophysics}
  \textbf{\bibinfo{volume}{14}}, \bibinfo{pages}{543} (\bibinfo{year}{2007}),
  ISSN \bibinfo{issn}{1023-5809},
  \urlprefix\url{http://www.nonlin-processes-geophys.net/14/543/2007/}.

\bibitem[{\citenamefont{Verheest}(2007{\natexlab{b}})}]{npg-14-545-2007}
\bibinfo{author}{\bibfnamefont{F.}~\bibnamefont{Verheest}},
  \bibinfo{journal}{Nonlinear Processes in Geophysics}
  \textbf{\bibinfo{volume}{14}}, \bibinfo{pages}{545}
  (\bibinfo{year}{2007}{\natexlab{b}}), ISSN \bibinfo{issn}{1023-5809},
  \urlprefix\url{http://www.nonlin-processes-geophys.net/14/545/2007/}.

\bibitem[{\citenamefont{McKenzie et~al.}(2005)\citenamefont{McKenzie, Dubinin,
  and Sauer}}]{npg-12-425-2005}
\bibinfo{author}{\bibfnamefont{J.~F.} \bibnamefont{McKenzie}},
  \bibinfo{author}{\bibfnamefont{E.~M.} \bibnamefont{Dubinin}},
  \bibnamefont{and} \bibinfo{author}{\bibfnamefont{K.}~\bibnamefont{Sauer}},
  \bibinfo{journal}{Nonlinear Processes in Geophysics}
  \textbf{\bibinfo{volume}{12}}, \bibinfo{pages}{425} (\bibinfo{year}{2005}),
  ISSN \bibinfo{issn}{1023-5809},
  \urlprefix\url{http://www.nonlin-processes-geophys.net/12/425/2005/}.

\bibitem[{\citenamefont{{Solomon} et~al.}(2006)\citenamefont{{Solomon},
  {Burrell}, {Andre}, {Baylor}, {Budny}, {Gohil}, {Groebner}, {Holcomb},
  {Houlberg}, and {Wade}}}]{solomonetal-pop-2006}
\bibinfo{author}{\bibfnamefont{W.~M.} \bibnamefont{{Solomon}}},
  \bibinfo{author}{\bibfnamefont{K.~H.} \bibnamefont{{Burrell}}},
  \bibinfo{author}{\bibfnamefont{R.}~\bibnamefont{{Andre}}},
  \bibinfo{author}{\bibfnamefont{L.~R.} \bibnamefont{{Baylor}}},
  \bibinfo{author}{\bibfnamefont{R.}~\bibnamefont{{Budny}}},
  \bibinfo{author}{\bibfnamefont{P.}~\bibnamefont{{Gohil}}},
  \bibinfo{author}{\bibfnamefont{R.~J.} \bibnamefont{{Groebner}}},
  \bibinfo{author}{\bibfnamefont{C.~T.} \bibnamefont{{Holcomb}}},
  \bibinfo{author}{\bibfnamefont{W.~A.} \bibnamefont{{Houlberg}}},
  \bibnamefont{and} \bibinfo{author}{\bibfnamefont{M.~R.}
  \bibnamefont{{Wade}}}, \bibinfo{journal}{Phys. Plasmas}
  \textbf{\bibinfo{volume}{13}} (\bibinfo{year}{2006}).

\bibitem[{\citenamefont{{Stacey}}(2006)}]{stacey-cpp06}
\bibinfo{author}{\bibfnamefont{W.~M.} \bibnamefont{{Stacey}}},
  \bibinfo{journal}{Contributions to Plasma Physics}
  \textbf{\bibinfo{volume}{46}}, \bibinfo{pages}{597} (\bibinfo{year}{2006}),
  \urlprefix\url{http://dx.doi.org/10.1002/ctpp.200610050}.

\bibitem[{\citenamefont{{Stacey} et~al.}(2006)\citenamefont{{Stacey},
  {Johnson}, and {Mandrekas}}}]{frc-pop-2006}
\bibinfo{author}{\bibfnamefont{W.~M.} \bibnamefont{{Stacey}}},
  \bibinfo{author}{\bibfnamefont{R.~W.} \bibnamefont{{Johnson}}},
  \bibnamefont{and}
  \bibinfo{author}{\bibfnamefont{J.}~\bibnamefont{{Mandrekas}}},
  \bibinfo{journal}{Phys. Plasmas} \textbf{\bibinfo{volume}{13}}
  (\bibinfo{year}{2006}),
  \urlprefix\url{http://link.aip.org/link/?PHPAEN/13/062508/1}.

\bibitem[{\citenamefont{Holcomb et~al.}(2006)\citenamefont{Holcomb, Makowski,
  Jayakumar, Allen, Ellis, Geer, Behne, Morris, Seppala, and
  Moller}}]{holcomb-10E506}
\bibinfo{author}{\bibfnamefont{C.~T.} \bibnamefont{Holcomb}},
  \bibinfo{author}{\bibfnamefont{M.~A.} \bibnamefont{Makowski}},
  \bibinfo{author}{\bibfnamefont{R.~J.} \bibnamefont{Jayakumar}},
  \bibinfo{author}{\bibfnamefont{S.~A.} \bibnamefont{Allen}},
  \bibinfo{author}{\bibfnamefont{R.~M.} \bibnamefont{Ellis}},
  \bibinfo{author}{\bibfnamefont{R.}~\bibnamefont{Geer}},
  \bibinfo{author}{\bibfnamefont{D.}~\bibnamefont{Behne}},
  \bibinfo{author}{\bibfnamefont{K.~L.} \bibnamefont{Morris}},
  \bibinfo{author}{\bibfnamefont{L.~G.} \bibnamefont{Seppala}},
  \bibnamefont{and} \bibinfo{author}{\bibfnamefont{J.~M.}
  \bibnamefont{Moller}}, \bibinfo{journal}{Rev. Sci. Instrum.}
  \textbf{\bibinfo{volume}{77}}, \bibinfo{pages}{10E506}
  (\bibinfo{year}{2006}),
  \urlprefix\url{http://link.aip.org/link/?RSI/77/10E506/1}.

\bibitem[{\citenamefont{Cortes et~al.}(2003)\citenamefont{Cortes, Hawkes,
  Lotte, Fenzi, Stratton, Hobirk, Angelis, Orsitto, Varandas, and to~the
  EFDA-JET~work program}}]{cortes-1596}
\bibinfo{author}{\bibfnamefont{S.~R.} \bibnamefont{Cortes}},
  \bibinfo{author}{\bibfnamefont{N.~C.} \bibnamefont{Hawkes}},
  \bibinfo{author}{\bibfnamefont{P.}~\bibnamefont{Lotte}},
  \bibinfo{author}{\bibfnamefont{C.}~\bibnamefont{Fenzi}},
  \bibinfo{author}{\bibfnamefont{B.~C.} \bibnamefont{Stratton}},
  \bibinfo{author}{\bibfnamefont{J.}~\bibnamefont{Hobirk}},
  \bibinfo{author}{\bibfnamefont{R.~D.} \bibnamefont{Angelis}},
  \bibinfo{author}{\bibfnamefont{F.}~\bibnamefont{Orsitto}},
  \bibinfo{author}{\bibfnamefont{C.~A.~F.} \bibnamefont{Varandas}},
  \bibnamefont{and} \bibinfo{author}{\bibfnamefont{C.}~\bibnamefont{to~the
  EFDA-JET~work program}}, \bibinfo{journal}{Rev. Sci. Instrum.}
  \textbf{\bibinfo{volume}{74}}, \bibinfo{pages}{1596} (\bibinfo{year}{2003}),
  \urlprefix\url{http://link.aip.org/link/?RSI/74/1596/1}.

\bibitem[{\citenamefont{Simmons}(1991)}]{simmons-91}
\bibinfo{author}{\bibfnamefont{G.~F.} \bibnamefont{Simmons}},
  \emph{\bibinfo{title}{Differential Equations with Applications and Historical
  Notes}} (\bibinfo{publisher}{McGraw-Hill, Inc.}, \bibinfo{address}{New York,
  NY}, \bibinfo{year}{1991}), \bibinfo{edition}{2nd} ed.

\bibitem[{\citenamefont{{Flanigan}}(1972)}]{flanigan}
\bibinfo{author}{\bibfnamefont{F.~J.} \bibnamefont{{Flanigan}}},
  \emph{\bibinfo{title}{Complex Variables: Harmonic and Analytic Functions}}
  (\bibinfo{publisher}{Aliyn and Bacon, Inc., Boston, MA},
  \bibinfo{year}{1972}).

\bibitem[{\citenamefont{{Binns} et~al.}(1992)\citenamefont{{Binns},
  {Lawrenson}, and {Trowbridge}}}]{BLT-92}
\bibinfo{author}{\bibfnamefont{K.~J.} \bibnamefont{{Binns}}},
  \bibinfo{author}{\bibfnamefont{P.~J.} \bibnamefont{{Lawrenson}}},
  \bibnamefont{and} \bibinfo{author}{\bibfnamefont{C.~W.}
  \bibnamefont{{Trowbridge}}}, \emph{\bibinfo{title}{The Analytical and
  Numerical Solution of Electric and Magnetic Fields}}
  (\bibinfo{publisher}{John Wiley and Sons}, \bibinfo{year}{1992}).

\bibitem[{\citenamefont{{Weisstein}}(2008)}]{wolf-lapeqn}
\bibinfo{author}{\bibfnamefont{E.~W.} \bibnamefont{{Weisstein}}},
  \emph{\bibinfo{title}{Laplace's equation}}, \bibinfo{howpublished}{From
  MathWorld--A Wolfram Web Resource} (\bibinfo{year}{2008}),
  \urlprefix\url{http://mathworld.wolfram.com/LaplacesEquation.html}.

\bibitem[{\citenamefont{{Chu}}(2007)}]{mingagrees}
\bibinfo{author}{\bibfnamefont{M.~S.} \bibnamefont{{Chu}}},
  \emph{\bibinfo{title}{private communication}} (\bibinfo{year}{2007}).

\bibitem[{\citenamefont{{Woods}}(2006)}]{woodsbook-06}
\bibinfo{author}{\bibfnamefont{L.~C.} \bibnamefont{{Woods}}},
  \emph{\bibinfo{title}{Theory of Tokamak Transport: New Aspects for Nuclear
  Fusion Reactor Design}} (\bibinfo{publisher}{Wiley-VCH},
  \bibinfo{year}{2006}).

\bibitem[{\citenamefont{{Tee} and {Wake}}(2007)}]{woodsobit}
\bibinfo{author}{\bibfnamefont{G.}~\bibnamefont{{Tee}}} \bibnamefont{and}
  \bibinfo{author}{\bibfnamefont{G.}~\bibnamefont{{Wake}}},
  \emph{\bibinfo{title}{Obituary: {Leslie Woods}}}, \bibinfo{howpublished}{The
  Guardian} (\bibinfo{year}{2007}).

\bibitem[{\citenamefont{Meserve et~al.}(1991)\citenamefont{Meserve, Bernstein,
  Frieman, Grunder, Hanflin, Koonin, Papay, Prager, Ripin, and
  Sessoms}}]{jfe18-85}
\bibinfo{author}{\bibfnamefont{R.~A.} \bibnamefont{Meserve}},
  \bibinfo{author}{\bibfnamefont{I.}~\bibnamefont{Bernstein}},
  \bibinfo{author}{\bibfnamefont{E.}~\bibnamefont{Frieman}},
  \bibinfo{author}{\bibfnamefont{H.}~\bibnamefont{Grunder}},
  \bibinfo{author}{\bibfnamefont{R.}~\bibnamefont{Hanflin}},
  \bibinfo{author}{\bibfnamefont{S.}~\bibnamefont{Koonin}},
  \bibinfo{author}{\bibfnamefont{L.}~\bibnamefont{Papay}},
  \bibinfo{author}{\bibfnamefont{S.}~\bibnamefont{Prager}},
  \bibinfo{author}{\bibfnamefont{B.}~\bibnamefont{Ripin}}, \bibnamefont{and}
  \bibinfo{author}{\bibfnamefont{A.}~\bibnamefont{Sessoms}},
  \bibinfo{journal}{Journal of Fusion Energy} \textbf{\bibinfo{volume}{18}},
  \bibinfo{pages}{85} (\bibinfo{year}{1991}),
  \urlprefix\url{http://www.springerlink.com/content/v5h28773621300rq}.

\end{thebibliography}
\end{document}